\documentclass[12pt, centerh1]{article}
\usepackage{amssymb,amsmath,amsthm,colonequals,color,multirow,graphicx,mathrsfs}
\usepackage{natbib}
\usepackage{setspace}

\theoremstyle{plain}

\theoremstyle{definition}

\theoremstyle{remark}

\newcommand{\btheta}{{\boldsymbol{\theta}}}
\newcommand{\bvtheta}{{\boldsymbol{\vartheta}}}

\newcommand\numberthis{\addtocounter{equation}{1}\tag{\theequation}}

\newcommand{\vecmu}{\mbox{\boldmath$\mu$}}

\newcommand{\vecx}{\mathbf{x}}

\newcommand{\vecX}{\mathbf{X}}

\newcommand{\vecV}{\mathbf{V}}

\newcommand{\matsig}{\mathbf\Sigma}
\newcommand{\matPsi}{\mathbf\Psi}

\newcommand{\tr}{\,\mbox{tr}}

\newcommand{\vecA}{\mathbf{A}}

\newcommand{\matm}{\mathbf{M}}

\newcommand{\vecalp}{\boldsymbol{\alpha}}
\newcommand{\vecc}{\text{vec}}
\newcommand{\fX}{\mathscr{X}}
\newcommand{\fV}{\mathscr{V}}

\date{}
\begin{document}
\title{A Matrix Variate Skew-$t$ Distribution}
\author{Michael P.B.\ Gallaugher and Paul D.\ McNicholas}
\date{\small Dept.\ of Mathematics \& Statistics, McMaster University, Hamilton, Ontario, Canada.}

\maketitle

\begin{abstract} 
Although there is ample work in the literature dealing with skewness in the multivariate setting, there is a relative paucity of work in the matrix variate paradigm. Such work is, for example, useful for modelling three-way data. A matrix variate skew-$t$ distribution is derived based on a mean-variance matrix normal mixture. An expectation-conditional maximization algorithm is developed for parameter estimation. Simulated data are used for illustration.\\[+8pt]
\end{abstract}

\section{Introduction}
Matrix variate distributions have proven to be useful for modelling three-way data, such as multivariate longitudinal data. However, in most cases, the underlying distribution has been elliptical such as the matrix variate normal and the matrix variate $t$ distributions. However, there has been relatively little work done on matrix variate data that can account for skewness present in the data. The work that has been carried out in the area of matrix variate skew distributions is mostly limited to the matrix variate skew-normal distribution. Herein, we derive a matrix variate skew-$t$ distribution. The remainder of this paper is laid out as follows. In Section~2, some background is presented. In Section~3, the density of the matrix variate skew-$t$ distribution is derived and a parameter estimation procedure is given. Section~4 looks at some simulations, and we conclude with a summary and some future work (Section~5).

\section{Background}

\subsection{Matrix Variate Distributions}
One natural method to model three-way data is to use a matrix-variate distribution. There are many examples in the literature of such distributions, the most well-known being the matrix-normal distribution. For notional clarity, we use $\vecX$ to denote a realization of a random matrix $\fX$. 
An $n\times p$ random matrix $\fX$ follows a matrix variate normal distribution with location parameter $\matm$ and scale matrices $\matsig$ and $\matPsi$ of dimensions $n\times n$ and $p\times p$, respectively. We write $\fX\sim\mathcal{N}_{n\times p}(\matm, \matsig, \matPsi)$ to denote such a random matrix and the density of $\fX$ can be written
\begin{equation}
f(\vecX~|~\matm, \matsig, \matPsi)=\frac{1}{(2\pi)^{\frac{np}{2}}|\matsig|^{\frac{p}{2}}|\matPsi|^{\frac{n}{2}}}\exp\left\{-\frac{1}{2}\tr\left(\matsig^{-1}(\vecX-\matm)\matPsi^{-1}(\vecX-\matm)'\right)\right\}.
\end{equation}

One well known property of the matrix variate normal distribution \citep{harrar08} is 
\begin{equation}
\fX\sim \mathcal{N}_{n\times p}(\matm,\matsig,\matPsi) \iff \vecc(\fX)\sim \mathcal{N}_{np}(\vecc(\matm),\matPsi\otimes \matsig )
\label{eq:normprop}
\end{equation}
where $\mathcal{N}_{np}(\cdot)$ is the multivariate normal density with dimension $np$, $\vecc(\matm)$ is the vectorization of $\matm$, and $\otimes$ is the Kronecker product.

Although the matrix variate normal is arguably the most mathematically tractable, there are examples of non-normal cases. One famous example is the Wishart distribution \citep{Wishart} arising as the distribution of the sample covariance matrix of a multivariate normal sample. More recently, however, there has been some work done in the area of matrix skew distributions such as the matrix-variate skew normal distribution, e.g., \cite{chen2005}, \cite{dominguez2007}, and \cite{harrar08}. More information on matrix variate distributions can be found in \cite{gupta99book}.
Very recently, there has also been work done in the area of finite mixtures. Specifically, \cite{Anderlucci15} looked at clustering and classification of multivariate longitudinal data using a mixture of matrix variate normal distributions. Also, \cite{dougru16}, looked at mixtures of matrix variate $t$ distributions.

\subsection{Normal Variance-Mean Mixtures}
Various multivariate distributions such as the multivariate $t$, and skew-$t$, the shifted asymmetric Laplace distribution, and the generalized hyperbolic distributions arise as special cases of a normal variance-mean mixture \citep[cf.][Ch.~6]{mcnicholas16a}.
In this formulation, the density of a $p$-dimensional random vector $\vecX$ takes the form
$$f(\vecx)=\int_{0}^{\infty}\phi_{p}(\vecx|\vecmu+w\vecalp,w\matsig)h(w|\btheta)dw,$$
which is equivalent to the representation
\begin{equation}
\vecX=\vecmu+W\vecalp+\sqrt{W}\vecV,
\label{eq:nmvmixture}
\end{equation}
where $\vecV\sim \mathcal{N}_p({\bf 0},\matsig)$ and $W>0$ is a latent random variable with density $h(w|\btheta)$.
The multivariate skew-$t$ distribution with $\nu$ degrees of freedom arises as a special case with $W\sim \text{IG}\left(\frac{\nu}{2},\frac{\nu}{2}\right)$, where  $\text{IG}(\cdot)$ denotes the inverse Gamma distribution with density function
$$
f(x|\alpha,\beta)=\frac{\beta^{\alpha}}{\Gamma(\alpha)}x^{-\alpha-1}\exp\left\{-\frac{\beta}{x}\right\}.
$$

\subsection{The Generalized Inverse Gaussian Distribution}
A random variable $Y$ has a generalized inverse Gaussian (GIG) distribution with parameters $a, b$ and $\lambda$ if its density function can be written as
$$
f(y|a, b, \lambda)=\frac{\left(\frac{a}{b}\right)^{\frac{\lambda}{2}}y^{\lambda-1}}{2K_{\lambda}(\sqrt{ab})}\exp\left\{-\frac{ay+\frac{b}{y}}{2}\right\},
$$
where
$$
K_{\lambda}(x)=\frac{1}{2}\int_{0}^{\infty}y^{\lambda-1}\exp\left\{-\frac{x}{2}\left(y+\frac{1}{y}\right)\right\}dy
$$
is the modified Bessel function of the third kind with index $\lambda$. Several functions of GIG random variables have tractable expected values, e.g.,
\begin{equation}
\mathbb{E}(Y)=\sqrt{\frac{b}{a}}\frac{K_{\lambda+1}(\sqrt{ab})}{K_{\lambda}(\sqrt{ab})},
\end{equation}
\begin{equation}
\mathbb{E}\left(\frac{1}{Y}\right)=\sqrt{\frac{a}{b}}\frac{K_{\lambda+1}(\sqrt{ab})}{K_{\lambda}(\sqrt{ab})}-\frac{2\lambda}{b},
\end{equation}
\begin{equation}
\mathbb{E}(\log Y)=\log\left(\sqrt{\frac{b}{a}}\right)+\frac{1}{K_{\lambda}(\sqrt{ab})}\frac{\partial}{\partial \lambda}K_{\lambda}(\sqrt{ab}),
\end{equation}
where $a,b\in\mathbb{R}^+$, $\lambda\in\mathbb{R}$, and $K_{\lambda}(\cdot)$ is the modified Bessel function of the third kind with index $\lambda$.
These results will prove to be useful for parameter estimation for the matrix-variate skew-$t$ distribution.

\section{Methodology}
\subsection{A Matrix Variate Skew-$t$ Distribution}
We will say that an $n\times p$ random matrix $\fX$ has a matrix variate skew-$t$ distribution, $\text{MVST}_{n\times p}(\matm,\vecA,\matsig,\matPsi,\nu)$, if $\fX$ can be written
\begin{equation}\label{eqn:that}
\fX=\matm+W\vecA+\sqrt{W}\fV,
\end{equation}
where $\matm$ and $\vecA$ are $n\times p$  matrices, $\fV \sim \mathcal{N}_{n \times p}\left(\bf{0}, \matsig , \matPsi \right)$, and $W\sim \text{IG}\left(\frac{\nu}{2},\frac{\nu}{2}\right)$. Analogous to its multivariate counterpart, $\matm$ is a location matrix, $\vecA$ is a skewness matrix, $\matsig$ and $\matPsi$ are scale matrices, and $\nu$ is the degrees of freedom.
It then follows that
$$
\fX|w\sim \mathcal{N}_{n\times p}\left(\matm+w\vecA,w\matsig,\matPsi\right)
$$
and thus the joint density of $\fX$ and $W$ is
\begin{align*}
f(\vecX,w~|~\bvtheta)&=f(\vecX~|~w)f(w)\\
&= \frac{\frac{\nu}{2}^{\frac{\nu}{2}}}{(2\pi)^{\frac{np}{2}}| \matsig |^{\frac{p}{2}} |\matPsi |^{\frac{n}{2}}\Gamma(\frac{\nu}{2})} w^{-\frac{\nu+np}{2}-1} \\
& \hspace{0.2in} \times \exp\left\{-\frac{1}{2w}\left[\tr\left(\matsig^{-1}(\vecX-\matm-w\vecA)\matPsi^{-1}(\vecX-\matm-w\vecA)'\right)+\nu\right]\right\}, \numberthis \label{eqn:joint}
\end{align*}
where $\bvtheta=(\matm,\vecA,\matsig,\matPsi,\nu)$.
We note that the exponential term in \eqref{eqn:joint} can be written as
$$
\exp\left\{\tr(\matsig^{-1}(\vecX-\matm)\matPsi^{-1}\vecA')\right\} \times \exp\left\{-\frac{1}{2}\left[\frac{\delta(\vecX;\matm,\matsig,\matPsi)+\nu}{w}+w\rho(\vecA,\matsig,\matPsi)\right]\right\},
$$
where
$$
\begin{array}{ccc}
\delta(\vecX;\matm,\matsig,\matPsi)=\tr(\matsig^{-1}(\vecX-\matm)\matPsi^{-1}(\vecX-\matm)') &\text{and}& \rho(\vecA,\matsig,\matPsi)=\tr(\matsig^{-1}\vecA\matPsi^{-1}\vecA').
\end{array}
$$
Therefore, the marginal density of $\fX$ is
\begin{align*}
f(\vecX)&=\int_{0}^{\infty}f(\vecX,w) dw\\
&=\frac{\frac{\nu}{2}^{\frac{\nu}{2}}}{(2\pi)^{\frac{np}{2}}| \matsig |^{\frac{p}{2}} |\matPsi |^{\frac{n}{2}}\Gamma(\frac{\nu}{2})}\exp\left\{\tr(\matsig^{-1}(\vecX-\matm)\matPsi^{-1}\vecA')\right\} \\
& \hspace{0.25in} \times \int_{0}^\infty w^{-\frac{\nu+np}{2}-1}\exp\left\{-\frac{1}{2}\left[\frac{\delta(\vecX;\matm,\matsig,\matPsi)+\nu}{w}+w\rho(\vecA,\matsig,\matPsi)\right]\right\}dw.
\end{align*}
Making the change of variables given by
$$
y=\frac{\sqrt{\rho(\vecA,\matsig,\matPsi)}}{\sqrt{\delta(X;\matm,\matsig,\matPsi)+\nu}}w
$$
we can write
\begin{align*}
f_{\text{MVST}}(\vecX~|~\bvtheta)=&\frac{2\left(\frac{\nu}{2}\right)^{\frac{\nu}{2}}\exp\left\{\tr(\matsig^{-1}(\vecX-\matm)\matPsi^{-1}\vecA') \right\} }{(2\pi)^{\frac{np}{2}}| \matsig |^{\frac{p}{2}} |\matPsi |^{\frac{n}{2}}\Gamma(\frac{\nu}{2})}  \left(\frac{\delta(\vecX;\matm,\matsig,\matPsi)+\nu}{\rho(\vecA,\matsig,\matPsi)}\right)^{-\frac{\nu+np}{4}} \\ & \qquad\qquad\qquad\qquad\times
 K_{-\frac{\nu+np}{2}}\left(\sqrt{\left[\rho(\vecA,\matsig,\matPsi)\right]\left[\delta(\vecX;\matm,\matsig,\matPsi)+\nu\right]}\right).
\end{align*}
The density of $\fX$, as derived here, is considered a matrix variate extension of the multivariate skew-$t$ density used by \cite{murray14b,murray14a}. 
As discussed by \cite{dutilleul99} and \cite{Anderlucci15} in the matrix variate normal case, the estimates of $\matsig$ and $\matPsi$ are unique only up to a multiplicative constant. Indeed, if we let $\tilde{\matsig}=v\matsig$ and $\tilde{\matPsi}=(1/v)\matPsi$, $v\ne0$, the likelihood is unchanged. This identifiability issue can be resolved, for example, by setting $\tr(\matsig)=n$ or $\tr(\matPsi)=p$. Note that $\tilde{\matPsi}\otimes\tilde{\matsig}=\matPsi\otimes\matsig$, so the estimate of the Kronecker product is unique. 

For the purposes of parameter estimation, note that the conditional density of $W$ is
\begin{align*}
f&(w~|~\vecX)=\frac{f(\vecX~|~w)f(w)}{f(\vecX)}\\
&=\frac{\left[{\rho(\vecA,\matsig,\matPsi)}/({\delta(\vecX;\matm,\matsig,\matPsi)+\nu})\right]^{\frac{\lambda}{2}}w^{\lambda-1}}{2K_{\lambda}(\sqrt{\rho(\vecA,\matsig,\matPsi)[\delta(\vecX;\matm,\matsig,\matPsi)+\nu}])}\exp\left\{-\frac{\rho(\vecA,\matsig,\matPsi)w+[{\delta(\vecX;\matm,\matsig,\matPsi)+\nu}]/{w}}{2}\right\}.
\end{align*}
Therefore, $$W~|~\vecX\sim GIG\left(\rho(\vecA,\matsig,\matPsi),\delta(\vecX;\matm,\matsig,\matPsi)+\nu,\lambda\right),$$ where $\lambda=-(\nu+np)/2$.

Finally, we note that
\begin{equation}\label{eqn:this}
\fX\sim \text{MVST}_{n\times p}(\matm,\vecA,\matsig,\matPsi,\nu) \iff \vecc(\fX) \sim \text{MST}_{np}(\vecc(\matm),\vecc(\vecA),\matPsi\otimes\matsig,\nu),
\end{equation}
where $\text{MST}_{np}(\cdot)$ denotes the multivariate skew-$t$ distribution with location parameter $\vecc(\matm)$, skewness parameter $\vecc(\vecA)$, scale matrix $\matPsi\otimes\matsig$, and $\nu$ degrees of freedom. This can be easily seen from the representation given in \eqref{eq:nmvmixture} and the property of the matrix normal distribution given in \eqref{eq:normprop}. 
Note that the normal variance-mean mixture representation \eqref{eqn:that} as well as the relationship with the multivariate skew-$t$ distribution \eqref{eqn:this} present two convenient methods to generate random matrices from the matrix variate skew $t$ distribution. The former is used in Section~4.

\subsection{Parameter Estimation}
Suppose we observe a sample of $N$ matrices $\vecX_1, \vecX_2, \ldots \vecX_N$ from an $n\times p$ matrix variate skew-$t$ distribution. As with the multivariate skew-$t$ distribution, we proceed as if the observed data is incomplete, and introduce the latent variables $W_1,\ldots,W_n$. The complete-data log-likelihood is then
\begin{align*}
\ell_{c}(\bvtheta)=&C+\frac{N\nu}{2}\log\left(\frac{\nu}{2}\right)-\frac{Np}{2}\log|\matsig|-\frac{Nn}{2}\log|\matPsi|-N\log\left(\Gamma\left(\frac{\nu}{2}\right)\right)-\frac{\nu}{2}\sum_{i=1}^N\log(w_i)\\ 
&+\frac{1}{2}\sum_{i=1}^N \tr \left(\matsig^{-1}(\vecX_i-\matm)\matPsi^{-1}\vecA'\right)+\frac{1}{2}\sum_{i=1}^N \tr \left(\matsig^{-1}\vecA\matPsi^{-1}(\vecX_i-\matm)'\right)
\\& -\frac{1}{2}\sum_{i=1}^N\frac{1}{w_i}\left[\tr(\matsig^{-1}(\vecX_i-\matm)\matPsi^{-1}(\vecX_i-\matm)')+\nu\right]-\frac{1}{2}\sum_{i=1}^Nw_i\tr(\matsig^{-1}\vecA\matPsi^{-1}\vecA'),
\end{align*}
where $C$ is constant with respect to the parameters.
We proceed by using an expectation-conditional maximization (ECM) algorithm \citep{meng93} described overleaf.
\newpage

\noindent {\bf 1) Initialization}: Initialize the parameters $\matm,\vecA,\matsig,\matPsi,\nu$. Set $t=0$

\noindent {\bf 2) E Step}: Update $a_i, b_i, c_i$, where
\begin{align*}
a_i^{(t+1)}&=\mathbb{E}(W_i|\vecX_i,\hat{\bvtheta}^{(t)})\\
&=\sqrt{\frac{\delta(\vecX_i;\hat{\matm}^{(t)},\hat{\matsig}^{(t)},\hat{\matPsi}^{(t)})+\hat{\nu}^{(t)}}{\rho(\hat{\vecA}^{(t)},\hat{\matsig}^{(t)},\hat{\matPsi}^{(t)})}}\frac{K_{\lambda^{(t)}+1}(\kappa^{(t)})}{K_{\lambda^{(t)}}(\kappa^{(t)})} \numberthis \label{eq:aup}\\
b_i^{(t+1)}&=\mathbb{E}\left(\frac{1}{W_i}\left.\right|\vecX_i,\hat{\bvtheta}^{(t)}\right)\\
&=\sqrt{\frac{\rho(\hat{\vecA}^{(t)},\hat{\matsig}^{(t)},\hat{\matPsi}^{(t)})}{\delta(\vecX_i;\hat{\matm}^{(t)},\hat{\matsig}^{(t)},\hat{\matPsi}^{(t)})+\hat{\nu}^{(t)}}}\frac{K_{\lambda^{(t)}+1}(\kappa^{(t)})}{K_{\lambda^{(t)}}(\kappa^{(t)})}+\frac{\hat{\nu}^{(t)}+np}{\delta(\vecX_i;\hat{\matm}^{(t)},\hat{\matsig}^{(t)},\hat{\matPsi}^{(t)})+\nu^{(t)}} \numberthis \label{eq:bup}\\
c_i^{(t+1)}&=\mathbb{E}(\log(W_i)|\vecX_i,\hat{\bvtheta}^{(t)})\\
&=\log\left(\sqrt{\frac{\delta(\vecX_i;\hat{\matm}^{(t)},\hat{\matsig}^{(t)},\hat{\matPsi}^{(t)})+\hat{\nu}^{(t)}}{\rho(\hat{\vecA}^{(t)},\hat{\matsig}^{(t)},\hat{\matPsi}^{(t)})}}\right)+\frac{1}{K_{\lambda^{(t)}}(\kappa^{(t)})} \numberthis \label{eq:cup} \left.\frac{\partial}{\partial \lambda}K_{\lambda}(\kappa^{(t)})\right|_{\lambda=\lambda^{(t)}}
\end{align*}
where 
$$
\kappa^{(t)}=\sqrt{[\rho(\hat{\vecA}^{(t)},\hat{\matsig}^{(t)},\hat{\matPsi}^{(t)})][\delta(\vecX_i;\hat{\matm}^{(t)},\hat{\matsig}^{(t)},\hat{\matPsi}^{(t)})+\hat{\nu}^{(t)}]},
$$
and 
$$
\lambda^{(t)}=-\frac{\nu^{(t)}+np}{2}
$$

\noindent {\bf 3) First CM Step}: Update the parameters $\matm,\vecA,\nu$. 
\begin{align*}
\hat{\matm}^{(t+1)}&=\frac{\sum_{i=1}^N\vecX_i\left(\overline{a}^{(t+1)}b^{(t+1)}_i-1\right)}{\sum_{i=1}^N\overline{a}^{(t+1)}b_i^{(t+1)}-N}, \numberthis \label{eq:muup}\\
\\
\hat{\vecA}^{(t+1)}&=\frac{\sum_{i=1}^N\vecX_i\left(\overline{b}^{(t+1)}-b^{(t+1)}_i\right)}{\sum_{i=1}^N\overline{a}^{(t+1)}b_i^{(t+1)}-N} \numberthis \label{eq:Aup},
\end{align*}where 
$\overline{a}^{(t+1)}=({1}/{N})\sum_{i=1}^Na_{i}^{(t+1)}$ and $\overline{b}^{(t+1)}=({1}/{N})\sum_{i=1}^Nb_{i}^{(t+1)}$.

The update for the degrees of freedom cannot be obtained in closed form. Instead we solve \eqref{eq:nuup} for $\nu$ to obtain $\hat{\nu}^{(t+1)}$. 
\begin{equation}
\log\left(\frac{\nu}{2}\right)+1-\varphi\left(\frac{\nu}{2}\right)-\frac{1}{N}\sum_{i=1}^N(b^{(t+1)}_i+c^{(t+1)}_i)=0,
\label{eq:nuup}
\end{equation}
where $\varphi(\cdot)$ is the digamma function.

\noindent {\bf 4) Second CM Step}: Update $\matsig$
\begin{equation}
\begin{split}
&\hat{\matsig}^{(t+1)}=\frac{1}{Np}\left[\sum_{i=1}^N\left(b^{(t+1)}_i\left(\vecX_i-\hat{\matm}^{(t+1)}\right)\left.\hat{\matPsi}^{(t)}\right.^{{-1}}\left(\vecX_i-\hat{\matm}^{(t+1)}\right)'\right.\right.\\&\left.\left.-\left.\hat{\vecA}^{(t+1)}\right.\left.\hat{\matPsi}^{(t)}\right.^{{-1}}\left(\vecX_i-\hat{\matm}^{(t+1)}\right)'
-\left(\vecX_i-\hat{\matm}^{(t+1)}\right)\left.\hat{\matPsi}^{(t)}\right.^{{-1}}\left.\hat{\vecA}^{(t+1)}\right.'\right.\right.\\&+\left.\left.a_i^{(t+1)}\left.\hat{\vecA}^{(t+1)}\right.\left.\hat{\matPsi}^{(t)}\right.^{{-1}}\left.\hat{\vecA}^{(t+1)}\right.'\right) \right]
\end{split}
\end{equation}

\noindent {\bf 5) Third CM Step}: Update $\matPsi$

\begin{equation}
\begin{split}
&\hat{\matPsi}^{(t+1)}=\frac{1}{Nn}\left[\sum_{i=1}^N\left(b^{(t+1)}_i\left(\vecX_i-\hat{\matm}^{(t+1)}\right)'\left.\hat{\matsig}^{(t+1)}\right.^{{-1}}\left(\vecX_i-\hat{\matm}^{(t+1)}\right)\right.\right.\\&\left.\left.-\left.\hat{\vecA}^{(t+1)}\right.'\left.\hat{\matsig}^{(t+1)}\right.^{{-1}}\left(\vecX_i-\hat{\matm}^{(t+1)}\right)
-\left(\vecX_i-\hat{\matm}^{(t+1)}\right)'\left.\hat{\matsig}^{(t+1)}\right.^{{-1}}\left.\hat{\vecA}^{(t+1)}\right.\right.\right.\\&+\left.\left.a_i^{(t+1)}\left.\hat{\vecA}^{(t+1)}\right.'\left.\hat{\matsig}^{(t+1)}\right.^{{-1}}\left.\hat{\vecA}^{(t+1)}\right. \right)\right]
\end{split}
\end{equation}

\noindent {\bf 6) Check Convergence}: If not converged, set $t=t+1$ and return to step 2.

Note that there are several options for determining convergence of this ECM algorithm. In the simulations in Section~4,  a criterion based on the Aitken acceleration \citep{aitken26} is used. The Aitken acceleration at iteration $t$ is
\begin{equation}\label{eqn:aa}
a^{(t)} = \frac{l^{(t+1)}-l^{(t)}}{l^{(t)}-l^{(t-1)}},
\end{equation}
where $l^{(t)}$ is the (observed) log-likelihood at iteration $t$. The quantity in \eqref{eqn:aa} can be used to derive an asymptotic estimate (i.e., an estimate of the value after very many iterations) of the log-likelihood at iteration $t+1$, i.e., $$l_{\infty}^{(t+1)} = l^{(t)} + \frac{1}{1-a^{(t)}}(l^{(t+1)}-l^{(t)})$$ \citep[cf.][]{bohning94, lindsay95}. As in \cite{mcnicholas10a}, we stop our EM algorithms when $l_{\infty}^{(t+1)}-l^{(t)} < \epsilon$, provided this difference is positive.

As discussed in \cite{dutilleul99} and \cite{Anderlucci15} for parameter estimation in the matrix variate normal case, the estimates of $\matsig$ and $\matPsi$ are unique only up to a multiplicative constant. Indeed, if we let $\tilde{\matsig}=v\matsig$ and $\tilde{\matPsi}=(1/v)\matPsi$, $v\ne0$, the likelihood is unchanged. However, we notice that, $\tilde{\matPsi}\otimes\tilde{\matsig}=\matPsi\otimes\matsig$, so the estimate of the Kronecker product is unique. 

\section{Simulations}
We conduct two simulations to illustrate the estimation of the parameters. In both simulations, we take 50 different datasets of size 100, from a $3\times 4$ matrix skew-$t$ distribution. Also, in both simulations, 
$$
\begin{array}{ll}
\matsig=\left(
\begin{array}{lll}
1 & 0.5 & 0.1 \\ 
0.5 & 1 & 0.5 \\
 0.1 & 0.5 & 1 \\
\end{array}\right),&\qquad
\matPsi=\left(
\begin{array}{rrrr}
1 & -0.5 & 0.5 & 0.1 \\ 
-0.5 & 1 & -0.5 & 0.6 \\
0.5 & -0.5 & 1 & -0.4 \\ 
0.1 & 0.6 & -0.4 & 1 \\
  \end{array}
\right),
\end{array}
$$
and $\nu=4$. In simulation 1, we took the location and skewness matrix to be $\matm_1$ and $\vecA_1$, respectively, and $\matm_2$ and $\vecA_2$ in simulation 2, where
$$
\begin{array}{ll}
\matm_1=\left(
\begin{array}{rrrr}
0& 1 & -1 & 0 \\ 
   1 & 0 & 0 & -1 \\ 
  0 & 1 & -1 & 0 \\ 
  \end{array}
\right),&\qquad
\vecA_1=\left(
\begin{array}{rrrr}
1 & -1 & 0 & 1 \\ 
1 & -1 & 0 & 1 \\
1 & -1 & 0 & 1 \\ 
  \end{array}
\right),\\
\\
\matm_2=\left(
\begin{array}{rrrr}
1 & -6 & -1 & -1 \\ 
  -3 & 5 & -4 & 1 \\ 
  1 & -4 & -1 & 5 \\
  \end{array}
\right),&\qquad
\vecA_2=\left(
\begin{array}{rrrr}
1 & -1 & 0.5 & 0 \\ 
  0.5 & -0.5 & 0.5 & 0.5 \\ 
  0 & 0 & 0.5 & 0 \\ 
  \end{array}
\right).
\end{array}
$$

\begin{figure}[!htb]
\centering
\includegraphics[width=0.6\textwidth,height=0.6\textwidth]{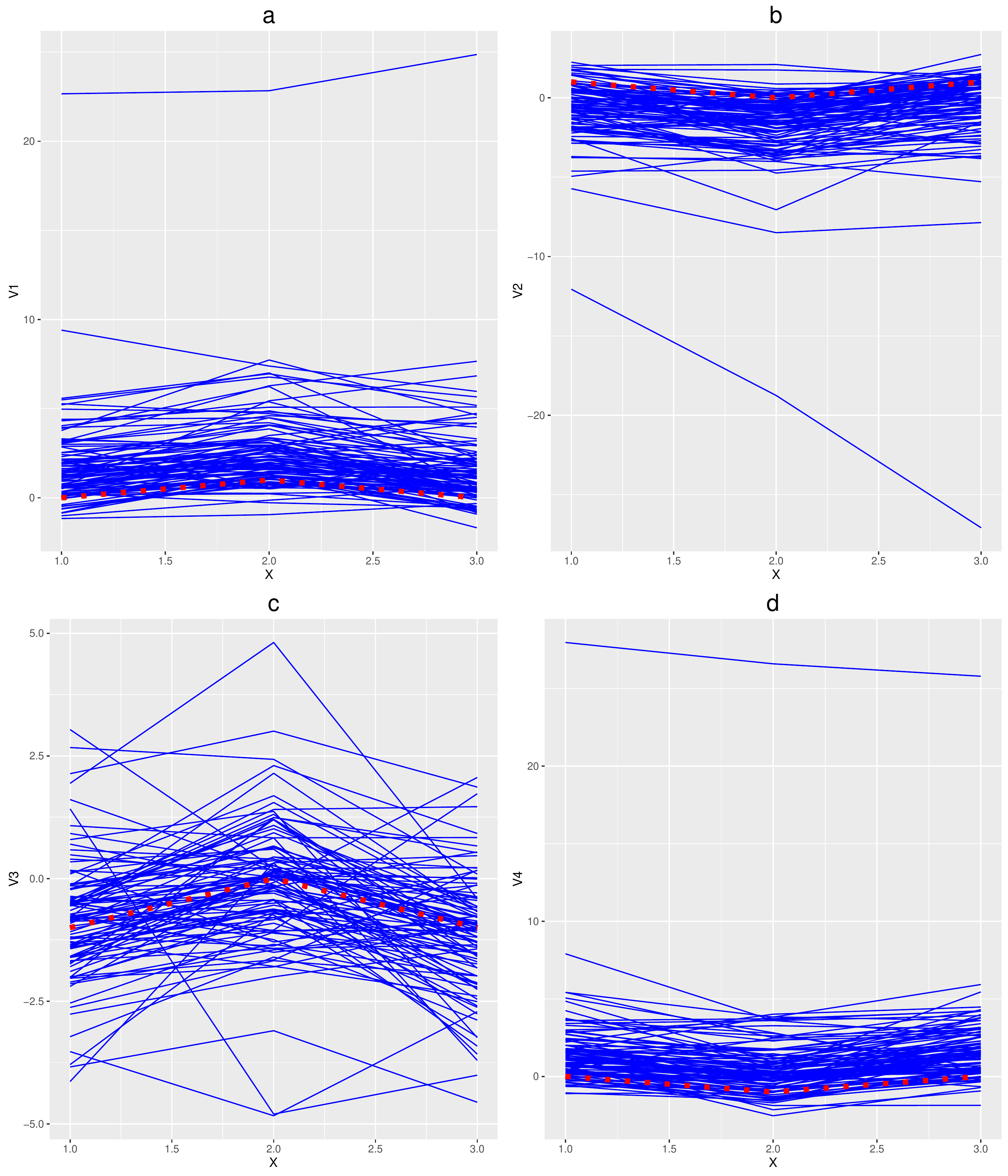}
\caption{Typical Marginals for Simulation 1 for (a) V1, (b) V2, (c) V3 and (d) V4. The red dashed lines denote the means.}
\label{fig:lineplots1}
\end{figure}

\begin{figure}[!htb]
\centering
\includegraphics[width=0.6\textwidth,height=0.6\textwidth]{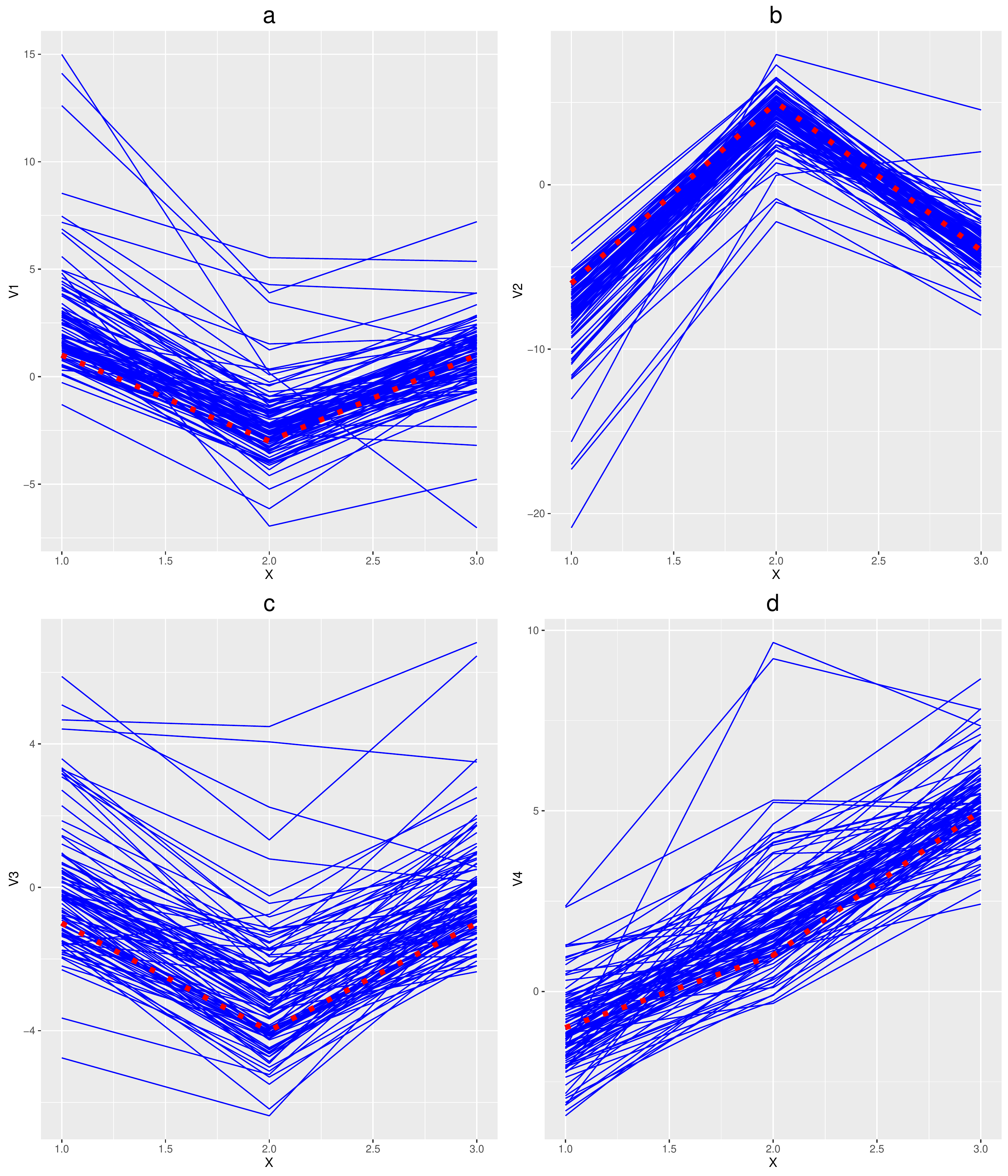}
\caption{Typical Marginals for Simulation 2 for (a) V1, (b) V2, (c) V3 and (d) V4. The red dashed lines denote the means.}
\label{fig:lineplots2}
\end{figure}

In Figures \ref{fig:lineplots1} and \ref{fig:lineplots2}, we show line plots of the marginals for each column (labelled V1, V2, V3, V4) of a typical dataset from simulations~1 and~2, respectively. The dashed red lines denote the means.
In Figure \ref{fig:lineplots1}, the skewness in columns 1, 2, and 4, for simulation 1, is very prominent when visually compared to column~3, which has zero skewness. The skewness is also apparent in the lineplots for simulation 2, however, because the values of the skewness are generally less than those for simulation 1, it is not as prominent.
The component-wise means of the parameters as well as the component wise standard deviations are given in Table~\ref{tab:results}. We see that the estimates of the mean matrix and skewness matrix are very close to the true value for both simulations. Moreover, we see that the estimates of $\matsig$ and $\matPsi$ also correspond approximately to the their true values, and thus so would the Kronecker product, which is not shown.
\begin{table}[!htb]
\centering
\caption{Component wise averages and standard deviations for the estimated parameters for simulations 1 and 2.}
\scalebox{0.55}{\begin{tabular}{cccccc}
\hline
Simulation& $\matm (sd)$&$\vecA (sd)$&$\matsig (sd)$&$\matPsi (sd)$&$\nu(sd)$\\
\hline
1 &
\shortstack{
$
\left[
\begin{array}{rrrr}
-0.04 & 1.04 & -1.01 & -0.02 \\ 
  1.01 & 0.03 & 0.03 & -1.01 \\ 
  0.01 & 1.04 & -0.97 & -0.01 \\ 
\end{array}
\right]
$
\\
$
\left(\left[
\begin{array}{rrrr}
0.212 & 0.176 & 0.175 & 0.176 \\ 
  0.181 & 0.216 & 0.158 & 0.151 \\ 
  0.185 & 0.206 & 0.137 & 0.146 \\ 
\end{array}
\right]\right)
$
}
&
\shortstack{
$
\left[
\begin{array}{rrrr}
1.07 & -1.06 & 0.03 & 1.04 \\ 
  1.01 & -1.04 & -0.01 & 1.03 \\ 
  1.02 & -1.03 & -0.01 & 1.04 \\ 
\end{array}
\right]
$
\\
$
\left(\left[
\begin{array}{rrrr}
0.197 & 0.174 & 0.120 & 0.180 \\ 
  0.177 & 0.192 & 0.113 & 0.167 \\ 
  0.182 & 0.201 & 0.088 & 0.169 \\ 
\end{array}
\right]\right)
$
}
&
\shortstack{
$
\left[
\begin{array}{rrr}
0.91 & 0.45 & 0.09 \\ 
  0.45 & 0.92 & 0.47 \\ 
  0.09 & 0.47 & 0.92 \\ 
\end{array}
\right]
$
\\
$
\left(\left[
\begin{array}{rrr}
0.111 & 0.074 & 0.055 \\ 
  0.074 & 0.108 & 0.069 \\ 
  0.055 & 0.069 & 0.096 \\ 
\end{array}
\right]\right)
$
}
&
\shortstack{
\\
\\
\\
$
\left[
\begin{array}{rrrr}
1.08 & -0.53 & 0.53 & 0.12 \\ 
  -0.53 & 1.06 & -0.53 & 0.64 \\ 
  0.53 & -0.53 & 1.08 & -0.44 \\ 
  0.12 & 0.64 & -0.44 & 1.09 \\ 
  \end{array}
\right]
$
\\
$
\left(\left[
\begin{array}{rrrr}
0.083 & 0.074 & 0.061 & 0.072 \\ 
  0.074 & 0.067 & 0.058 & 0.064 \\ 
  0.061 & 0.058 & 0.074 & 0.058 \\ 
  0.072 & 0.064 & 0.058 & 0.089 \\ 
  \end{array}
\right]\right)
$
}
& 
\shortstack{
4.22
\\
(0.63)
}
\\
\hline
2 &
\shortstack{
$
\left[
\begin{array}{rrrr}
0.99 & -6.01 & -0.99 & -1.02 \\ 
  -2.98 & 4.98 & -3.97 & 0.96 \\ 
  1.00 & -3.99 & -0.98 & 4.99 \\ 
\end{array}
\right]
$
\\
$
\left(\left[
\begin{array}{rrrr}
0.170 & 0.183 & 0.166 & 0.153 \\ 
  0.218 & 0.180 & 0.202 & 0.159 \\ 
  0.177 & 0.195 & 0.190 & 0.147 \\ 
\end{array}
\right]\right)
$
}
&
\shortstack{
$
\left[
\begin{array}{rrrr}
1.03 & -1.02 & 0.51 & 0.01 \\ 
  0.50 & -0.51 & 0.49 & 0.52 \\ 
  0.01 & -0.02 & 0.50 & 0.00 \\ 
\end{array}
\right]
$
\\
$
\left(\left[
\begin{array}{rrrr}
0.165 & 0.183 & 0.125 & 0.099 \\ 
  0.147 & 0.134 & 0.140 & 0.133 \\ 
  0.121 & 0.127 & 0.132 & 0.112 \\ 
\end{array}
\right]\right)
$
}
&
\shortstack{
$
\left[
\begin{array}{rrr}
0.92 & 0.45 & 0.08 \\ 
  0.45 & 0.91 & 0.45 \\ 
  0.08 & 0.45 & 0.92 \\ 
\end{array}
\right]
$
\\
$
\left(\left[
\begin{array}{rrr}
0.104 & 0.066 & 0.037 \\ 
  0.066 & 0.110 & 0.067 \\ 
  0.037 & 0.067 & 0.103 \\ 
\end{array}
\right]\right)
$
}
&
\shortstack{
\\
\\
\\
$
\left[
\begin{array}{rrrr}
1.09 & -0.54 & 0.54 & 0.11 \\ 
  -0.54 & 1.07 & -0.53 & 0.63 \\ 
  0.54 & -0.53 & 1.06 & -0.42 \\ 
  0.11 & 0.63 & -0.42 & 1.05 \\ 
  \end{array}
\right]
$
\\
$
\left(\left[
\begin{array}{rrrr}
0.087 & 0.068 & 0.064 & 0.068 \\ 
  0.068 & 0.083 & 0.055 & 0.079 \\ 
  0.064 & 0.055 & 0.066 & 0.055 \\ 
  0.068 & 0.079 & 0.055 & 0.090 \\ 
  \end{array}
\right]\right)
$
}
& 
\shortstack{
4.22
\\
(0.92)
}\\
\hline
\end{tabular}}
\label{tab:results}
\end{table}

\section{Discussion}
The density of a matrix variate skew-$t$ distribution was derived. This distribution can be considered as a three-way extension of the multivariate skew-$t$ distribution. Parameter estimation was carried out using an ECM algorithm. Because the formulation of multivariate skew-$t$ distribution this work is based on is a special case of the generalized hyperbolic distribution, it is reasonable to postulate an extension to a broader class of matrix variate distributions. Ongoing work considers a finite mixture of matrix variate skew-$t$ distributions for clustering and classification of three-way data. 

\subsection*{Acknowledgements}
The authors gratefully acknowledge the financial support provided by the Vanier Canada Graduate Scholarships (Gallaugher) and the Canada Research Chairs program (McNicholas).


\begin{thebibliography}{}

\bibitem[\protect\citeauthoryear{Aitken}{Aitken}{1926}]{aitken26}
Aitken, A.~C. (1926).
\newblock A series formula for the roots of algebraic and transcendental
  equations.
\newblock {\em Proceedings of the Royal Society of Edinburgh\/}~{\em 45},
  14--22.

\bibitem[\protect\citeauthoryear{Anderlucci, Viroli, et~al.}{Anderlucci
  et~al.}{2015}]{Anderlucci15}
Anderlucci, L., C.~Viroli, et~al. (2015).
\newblock Covariance pattern mixture models for the analysis of multivariate
  heterogeneous longitudinal data.
\newblock {\em The Annals of Applied Statistics\/}~{\em 9\/}(2), 777--800.

\bibitem[\protect\citeauthoryear{B\"{o}hning, Dietz, Schaub, Schlattmann, and
  Lindsay}{B\"{o}hning et~al.}{1994}]{bohning94}
B\"{o}hning, D., E.~Dietz, R.~Schaub, P.~Schlattmann, and B.~Lindsay (1994).
\newblock The distribution of the likelihood ratio for mixtures of densities
  from the one-parameter exponential family.
\newblock {\em Annals of the Institute of Statistical Mathematics\/}~{\em 46},
  373--388.

\bibitem[\protect\citeauthoryear{Chen and Gupta}{Chen and
  Gupta}{2005}]{chen2005}
Chen, J.~T. and A.~K. Gupta (2005).
\newblock Matrix variate skew normal distributions.
\newblock {\em Statistics\/}~{\em 39\/}(3), 247--253.

\bibitem[\protect\citeauthoryear{Do{\u{g}}ru, Bulut, and Arslan}{Do{\u{g}}ru
  et~al.}{2016}]{dougru16}
Do{\u{g}}ru, F.~Z., Y.~M. Bulut, and O.~Arslan (2016).
\newblock Finite mixtures of matrix variate t distributions.
\newblock {\em Gazi University Journal of Science\/}~{\em 29\/}(2), 335--341.

\bibitem[\protect\citeauthoryear{Dom{\'\i}nguez-Molina,
  Gonz{\'a}lez-Far{\'\i}as, Ramos-Quiroga, and Gupta}{Dom{\'\i}nguez-Molina
  et~al.}{2007}]{dominguez2007}
Dom{\'\i}nguez-Molina, J.~A., G.~Gonz{\'a}lez-Far{\'\i}as, R.~Ramos-Quiroga,
  and A.~K. Gupta (2007).
\newblock A matrix variate closed skew-normal distribution with applications to
  stochastic frontier analysis.
\newblock {\em Communications in Statistics--Theory and Methods\/}~{\em
  36\/}(9), 1691--1703.

\bibitem[\protect\citeauthoryear{Dutilleul}{Dutilleul}{1999}]{dutilleul99}
Dutilleul, P. (1999).
\newblock The MLE algorithm for the matrix normal distribution.
\newblock {\em Journal of Statistical Computation and Simulation\/}~{\em
  64\/}(2), 105--123.

\bibitem[\protect\citeauthoryear{Gupta and Nagar}{Gupta and
  Nagar}{1999}]{gupta99book}
Gupta, A.~K. and D.~K. Nagar (1999).
\newblock {\em Matrix Variate Distributions}.
\newblock Boca Raton: Chapman \& Hall/CRC Press.

\bibitem[\protect\citeauthoryear{Harrar and Gupta}{Harrar and
  Gupta}{2008}]{harrar08}
Harrar, S.~W. and A.~K. Gupta (2008).
\newblock On matrix variate skew-normal distributions.
\newblock {\em Statistics\/}~{\em 42\/}(2), 179--194.

\bibitem[\protect\citeauthoryear{Lindsay}{Lindsay}{1995}]{lindsay95}
Lindsay, B.~G. (1995).
\newblock Mixture models: Theory, geometry and applications.
\newblock In {\em NSF-CBMS Regional Conference Series in Probability and
  Statistics}, Volume~5. California: Institute of Mathematical Statistics:
  Hayward.

\bibitem[\protect\citeauthoryear{McNicholas}{McNicholas}{2016}]{mcnicholas16a}
McNicholas, P.~D. (2016).
\newblock {\em Mixture Model-Based Classification}.
\newblock Boca Raton: Chapman \& Hall/CRC Press.

\bibitem[\protect\citeauthoryear{McNicholas, Murphy, McDaid, and
  Frost}{McNicholas et~al.}{2010}]{mcnicholas10a}
McNicholas, P.~D., T.~B. Murphy, A.~F. McDaid, and D.~Frost (2010).
\newblock Serial and parallel implementations of model-based clustering via
  parsimonious {G}aussian mixture models.
\newblock {\em Computational Statistics and Data Analysis\/}~{\em 54\/}(3),
  711--723.

\bibitem[\protect\citeauthoryear{Meng and Rubin}{Meng and Rubin}{1993}]{meng93}
Meng, X.-L. and D.~B. Rubin (1993).
\newblock Maximum likelihood estimation via the {ECM} algorithm: a general
  framework.
\newblock {\em Biometrika\/}~{\em 80}, 267--278.

\bibitem[\protect\citeauthoryear{Murray, Browne, and McNicholas}{Murray
  et~al.}{2014a}]{murray14b}
Murray, P.~M., R.~B. Browne, and P.~D. McNicholas (2014a).
\newblock Mixtures of skew-t factor analyzers.
\newblock {\em Computational Statistics and Data Analysis\/}~{\em 77},
  326--335.

\bibitem[\protect\citeauthoryear{Murray, McNicholas, and Browne}{Murray
  et~al.}{2014b}]{murray14a}
Murray, P.~M., P.~D. McNicholas, and R.~B. Browne (2014b).
\newblock A mixture of common skew-$t$ factor analyzers.
\newblock {\em Stat\/}~{\em 3\/}(1), 68--82.

\bibitem[\protect\citeauthoryear{Wishart}{Wishart}{1928}]{Wishart}
Wishart, J. (1928).
\newblock The generalised product moment distribution in samples from a normal
  multivariate population.
\newblock {\em Biometrika\/}, 32--52.

\end{thebibliography}
 \end{document}